\documentclass[prl,twocolumn,showpacs,floatfix,amsmath,amssymb,superscriptaddress]{revtex4-2}
\usepackage{amsfonts}
\usepackage{stmaryrd}
\usepackage{ulem}
\usepackage{bbm}
\usepackage{mathrsfs}
\usepackage{tipa}
\usepackage{amssymb}
\usepackage{txfonts}
\usepackage{graphicx}
\usepackage{dcolumn}
\usepackage{epstopdf}
\usepackage[colorlinks,linkcolor=blue,urlcolor=black,citecolor=blue]{hyperref}
\usepackage{multirow}
\usepackage{subfigure}
\usepackage{url}
\usepackage{upgreek}
\usepackage[utf8]{inputenc}
\usepackage[english]{babel}

\begin{document}

\newcommand*{\cm}{cm$^{-1}$\,}
\newcommand*{\Tc}{T$_c$\,}
\newcommand*{\CRO}{CaRuO$_3$\,}
\newcommand*{\SRO}{SrRuO$_3$\,}
\newcommand*{\SCRO}{Sr$_\text{1-x}$Ca$_\text{x}$RuO$_3$}

\title{Strong terahertz third-harmonic generation by kinetic heavy quasiparticles in \CRO}

\author{C.~Reinhoffer}
\affiliation{Institute of Physics II, University of Cologne, 50937 Cologne, Germany}

\author{Sven~Esser}
\affiliation{Experimental Physics VI, Center for Electronic Correlations and Magnetism, University of Augsburg, 86159 Augsburg, Germany}

\author{Sebastian~Esser}
\email{Present Address: Department of Applied Physics and Quantum-Phase Electronics Center (QPEC), The University of Tokyo, Bunkyo-ku, Tokyo 113-8656, Japan}
\affiliation{Experimental Physics VI, Center for Electronic Correlations and Magnetism, University of Augsburg, 86159 Augsburg, Germany}

\author{E.~A.~Mashkovich}
\author{S.~Germanskiy}
\affiliation{Institute of Physics II, University of Cologne, 50937 Cologne, Germany}

\author{P.~Gegenwart}
\affiliation{Experimental Physics VI, Center for Electronic Correlations and Magnetism, University of Augsburg, 86159 Augsburg, Germany}

\author{F. Anders}
\affiliation{Department of Physics, TU Dortmund University, 44227 Dortmund, Germany}

\author{P.~H.~M.~van~Loosdrecht}
\affiliation{Institute of Physics II, University of Cologne, 50937 Cologne, Germany}

\author{Zhe Wang}
\email{Corresponding author: zhe.wang@tu-dortmund.de}
\affiliation{Department of Physics, TU Dortmund University, 44227 Dortmund, Germany}
\affiliation{Institute of Physics II, University of Cologne, 50937 Cologne, Germany}

\date{\today}

\begin{abstract}
We report on time-resolved nonlinear terahertz spectroscopy of a strongly correlated ruthenate, \CRO, as a function of temperature, frequency and terahertz field strength. Third-harmonic radiation for frequencies up to 2.1~THz is observed evidently at low temperatures below 80~K, where the low-frequency linear dynamical response deviates from the Drude model and a coherent heavy quasiparticle band emerges by strong correlations associated with the Hund's coupling.
Phenomenologically, by taking an experimentally observed frequency-dependent scattering rate, the deviation of the field driven kinetics from the Drude behavior is reconciled in a time-dependent Boltzmann description, which allows an attribution of the observed third-harmonic generation to the terahertz field driven nonlinear kinetics of the heavy quasiparticles. 
\end{abstract}

\maketitle

Driven by quantum fluctuations associated with the Heisenberg's uncertainty principle~\cite{Coleman2005}, a phase transition can occur at absolute zero temperature through tuning of external parameters such as chemical substitution and applied pressure \cite{Loehneysen2007,Gegenwart2008}. A variety of novel physical phenomena are observed in the quantum critical metallic systems \cite{Coleman2005,Loehneysen2007,Gegenwart2008}. Whereas in an ordinary metal the electrons can be treated as a weakly interacting liquid of fermions (i.e. Fermi liquid), a metal close to the quantum critical point is more complex which results from competition of different interactions and exhibits various non-Fermi-liquid behaviors~\cite{Schofield1999,Loehneysen2007,Gegenwart2008}.

The 4\textit{d} transition-metal oxides are of particular interest in the investigation of exotic quantum phenomena induced by electron correlations \cite{Stewart01}. 
The 4\textit{d} orbitals are characterized by a sizable onsite Coulomb repulsion and at the same time more extended in space than their 3\textit{d} counterparts. Therefore, the 4\textit{d} transition-metal oxides are not necessarily correlation-induced Mott insulators, but can be a strongly correlated metal whose properties cannot be fully described by Fermi-liquid theory \cite{Berthod2013,Schneider2014}. With more than one electron (or hole) in the 4\textit{d} shell, the Hund's coupling between the electrons is important and may even play a dominant role over the effective Coulomb repulsion in determining the magnetic and transport properties~\cite{Dang2015a,Dang2015}.

Based on the Ru$^{4+}$ (4$d^4$) ions the perovskite ruthenates \textit{A}RuO$_3$ (\textit{A} = Ca or Sr) are representative examples of quantum critical metals.
Characterized by a tilt and rotation of each RuO$_6$ octahedron from the ideal cubic perovskite structure, the orthorhombic distortion in CaRuO$_3$ is slightly greater than in SrRuO$_3$.
The small structural difference already leads to very different physical properties, since the onsite Coulomb repulsion and the Hund's coupling is only fine balanced which is sensitive to weak perturbation \cite{Dang2015}.
Whereas SrRuO$_3$ exhibits ferromagnetism at low temperatures, CaRuO$_3$ is located in an adjacent paramagnetic phase and very close to the ferromagnetic quantum critical point \cite{Dang2015}.

Various unusual metallic properties are observed in CaRuO$_3$.
Above 1.5~K the temperature dependence of its dc electrical resistivity deviates clearly from the characteristic quadratic dependence for a Fermi liquid \cite{Klein1999,Khalifah2004,Cao2008,Schneider2010}.
The optical conductivity of CaRuO$_3$ does not simply follow the Drude model \cite{Lee2002,Kamal2006,Schneider2010,Geiger2016,Wang2021}. Above about 0.6~THz, the frequency dependent optical conductivity cannot be described by a Fermi liquid theory \cite{Berthod2013,Schneider2014}.
Below 100~K angle-resolved photoemission spectroscopy revealed a well-defined heavy quasiparticle band with an enhanced effective mass of 13.5$m_e$ \cite{Liu2018}, clearly indicating the effects of strong correlations in CaRuO$_3$~\citep{Dang2015,Rondinelli_Caffrey_Sanvito_Spaldin_2008, Kumar_Prajapati_Dagar_Vagadia_Rana_Tonouchi_2020}.

An established approach to characterize a quantum critical system is based on the featured divergent behavior of its thermodynamic quantities, such as thermal expansion, specific heat, or Grüneisen ratio, which exhibit universal scaling with zero temperature being approached \cite{Zhu03}.
Also the dynamical response functions, e.g. optical conductivity, can follow a universal dependence on frequency, which is characteristic for a quantum critical point, see e.g. \cite{Damle97,smakov05,Krempa12,Lucas17}.
However, beyond the thermal equilibrium these quantities may not be well defined.
In contrast, nonlinear transport properties have been predicted to be very sensitive to quantum phase transition and can exhibit characteristics for a quantum critical system \cite{Dalidovich04,Green05,Oka05,Mitra06,Hogen2008,Shao22}.
For instance, close to a magnetic quantum critical point of a metallic system the current density scales nonlinearly with the applied electric field strength \cite{Hogen2008};
In an insulating quantum critical system of the one-dimensional Hubbard model, very efficient high-harmonic radiation can be generated due to optically induced interband transition \cite{Shao22}.
Motivated by these theoretical studies, in this work we investigate the ultrafast nonlinear transport behavior of the proximate quantum critical metallic system CaRuO$_3$ by using time-resolved terahertz (THz) third-harmonic generation (THG) spectroscopy.

Comparing with the short electrical current pulses of microseconds which lead primarily to Joule heating effects in the sample \cite{Esser2014}, the THz pulses with much shorter pulse duration probe directly the sub-picosecond (ps) nonlinear transport properties without considerable heating issues.
The nonlinear current related to the THz field driven kinetics of the quasiparticles is studied by measurement of THz THG.
The third-harmonic radiation is observed in the frequency range where the dynamical response of the system is beyond a description by the Fermi-liquid theory \cite{Berthod2013,Schneider2014}.
In contrast to the mechanism involving interband transitions \cite{Shao22,Murakami22} which requires higher-energy pump pulses e.g. mid-infrared \cite{Uchida22}, in our experiment only the bands in the vicinity (a few meV) of the Fermi surface are essentially responsible for the THz harmonic generation.
Our results show that third-order nonlinear THz susceptibility in CaRuO$_3$ is resolvable below 80~K and increases evidently with decreasing temperature. For different driving frequencies, the normalized temperature-dependence curves overlap very well with each other. The master curve resembles the temperature dependent evolution of the spectral weight of the heavy quasiparticle band \cite{Liu2018}, pointing to the crucial role of the strong correlations to the observed nonlinear responses.

\begin{figure}[t]
\includegraphics[width=8cm]{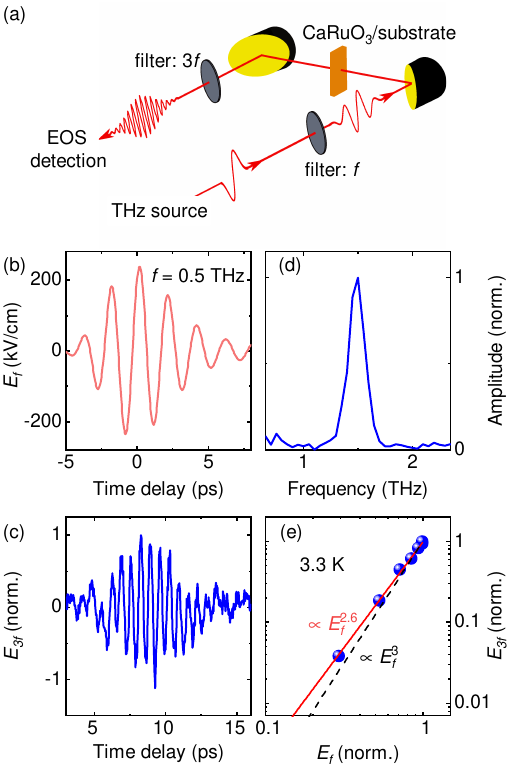}
\caption{\label{Fig1}  
(a) Illustration of the THz THG experiment. THG from CaRuO$_3$ thin film is measured in transmission configuration and the THz electric field is detected by electro-optic sampling (EOS).
(b) Electric field of the THz pump pulse with a central frequency of $f=0.5$ THz. 
(c) THz electric field emitted from \CRO at 3.3~K measured through a $3f$-bandpass (BP) filter. 
(d) Spectrum obtained by Fourier transformation of the waveform in panel (c). 
(e) THG amplitude $E_{3f}$ vs driving field amplitude $E_{f}$, which follows a power law dependence of $E_{3f} \propto E^{2.6}_{f}$ (solid line) slightly deviated from the perturbative dependence $\propto E^3_{f}$ (dashed line).}
\end{figure}

High-quality \CRO thin films were grown using metal-organic aerosol deposition technique \cite{Schneider2014} on 3° miscut (110) oriented NdGaO$_3$.
We carried out the THz spectroscopic measurements on a thin film with a residual resistivity ratio (RRR) of 35 and a thickness of $43.7$~nm which has been characterized previously \cite{Esser2014}.
A continuous helium-flow cryostat was installed for measurements at different temperatures.
Broadband THz radiation was generated in LiNbO$_3$ using the tilted-front optical rectification technique \cite{Hirori2011, Hebling2002} based on a Ti:sapphire amplified laser (800~nm, 1~kHz, 6~mJ). 
Narrow-band THz driving pulses with centre frequencies of $f=0.4$, 0.5, and 0.7~THz were prepared by using corresponding bandpass filters with full width at half maximum of $0.2f$ and an out-of-band transmission of $<-30$~dB [see Fig.~\ref{Fig1}(b) for $f=0.5$~THz].
The emitted THz electric field was recorded by electro-optical sampling (EOS) \cite{Wu1995} [see Fig.~\ref{Fig1}(a) for an illustration]. For the detection of THG another bandpass filter with a centre frequency of $3f$ was placed after the sample to suppress the linear response.
For fluence dependent measurements [Fig.~\ref{Fig1}(e)], two wire-grid polarizers were installed in front of the sample.

The crystal structure of CaRuO$_3$ belongs to the centrosymmetric \textit{Pbnm} space group, therefore the even-order nonlinear susceptibilities vanish and we will focus on the third-harmonic generation. 
For a driving pulse of $f=0.5$~THz, the emitted THz electric field from the \CRO thin film at 3.3~K is shown in Fig.~\ref{Fig1}(c). 
An evident oscillation of the electric field in time domain corresponding to a frequency of $3f=1.5$~THz is the third-harmonic generation, which can be directly read from its Fourier transform spectrum [see Fig.~\ref{Fig1}(d)].
Figure~\ref{Fig1}(e) displays fluence dependence of the integrated amplitude $E_{3f}$ of the emitted electric field,
which exhibits a power law behavior of $E_{3f} \propto E^{2.6}_f$ slightly deviated from the perturbative dependence of $\propto E^{3}_f$.

\begin{figure}[t]
\includegraphics[width=8.4cm]{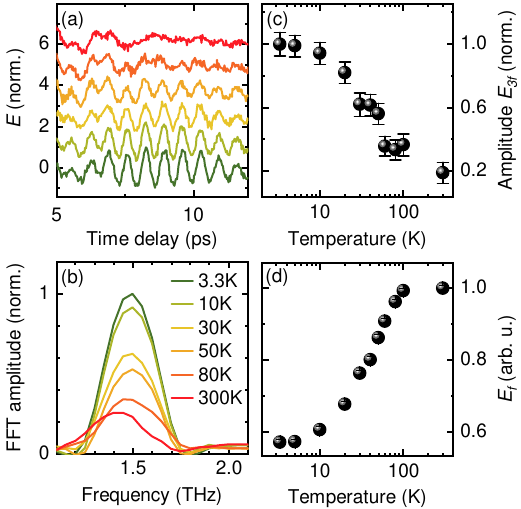}
\caption{\label{Fig2} 
(a) THG signal in time domain at different temperatures.
(b) Fourier amplitude of the signal in panel (a). 
(c) THG amplitude $E_{3f}$ as a function of temperature. 
(d) The transmitted amplitude $E_f$ of the fundamental frequency.}
\end{figure}

We further characterize the nonlinear THz response by measuring the THG as a function of temperature.
As shown in Fig.~\ref{Fig2}(a), the strongest THG signal is observed at 3.3~K in the time-delay window from 5 to 12.5~ps.
This signal decreases gradually with increasing temperature.
At 80~K the signal after 7~ps becomes nearly indiscernible, while before 7~ps a weak signal persists until 300~K.
At 300~K our fluence dependent measurement shows that this persisting signal follows a linear dependence on the fluence, thus is a transmission of the driving pulse rather than third-harmonic radiation. 
The enhancement of the signal after 7~ps with decreasing temperature is more clearly seen in the Fourier transform spectra in Fig.~\ref{Fig2}(b).
The integrated amplitude of the emitted $3f$ and $f$ components is presented in Fig.~\ref{Fig2}(c) and Fig.~\ref{Fig2}(d), respectively, as a function of temperature.
Below 80~K the substantial increase of the third-harmonic radiation [Fig.~\ref{Fig2}(c)] is accompanied with an evident drop of the transmission for the fundamental frequency $f=0.5$~THz [Fig.~\ref{Fig2}(d)].
The monotonic decrease of transmission at low frequencies reflects an enhanced metallic response with decreasing temperature, which is consistent with the observation of reduced dc resistivity and higher THz reflectivity~\cite{Klein1999,Schneider2014}.

\begin{figure}[t]
\includegraphics[width=8.6cm]{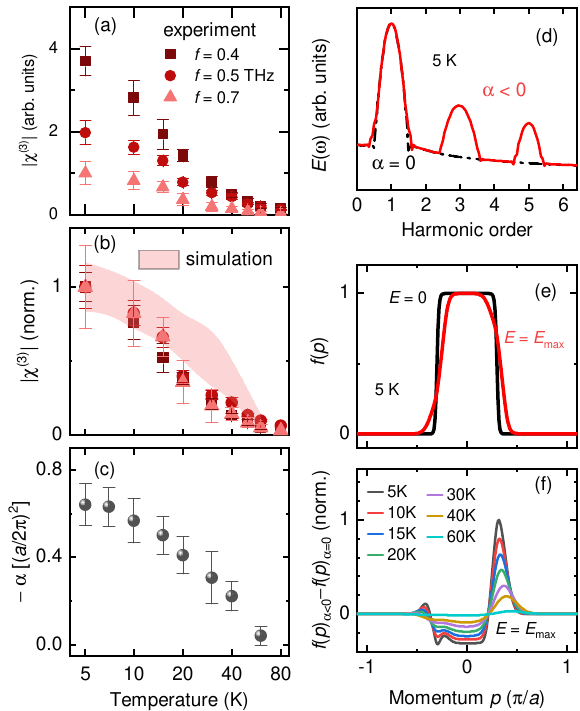}
\caption{\label{Fig3} 
(a) Experimentally determined $|\chi^{(3)}| \propto \frac{E_{3f}}{E_f^3}$ as a function of temperature for driving frequencies of 0.4, 0.5, and 0.7~THz.
(b) Normalized $|\chi^{(3)}|$ for THz drive frequencies of $f=0.4$, 0.5, and 0.7~THz follow the same temperature dependence.
The shaded area represents the simulated $|\chi^{(3)}|$ where the width reflects the uncertainties of experimentally determined $\alpha$ in (c) \cite{Schneider2014}.  
(d) Simulated spectral amplitude of emitted electric field for $\alpha = 0$ and $\alpha < 0$ at 5~K versus harmonic order. 
(e) Snapshots of charge-carrier distribution $f(p)$ at zero and peak THz electric field $E_{\text{max}}$ at 5~K.
(f) Changes of charge-carrier distribution at $E_{\text{max}}$ for different experimental values of $\alpha$ with respect to that of $\alpha = 0$.}
\end{figure}

For a fixed driving field, the observed temperature dependent transmission indicates that the effective THz field in the thin film varies with temperature, therefore we use a nonlinear susceptibility as defined by $|\chi^{(3)}| \propto \frac{E_{3f}}{E_f^3}$ to characterize the nonlinear dynamical responses of CaRuO$_3$ at different temperatures, also for different THz field strengths and frequencies. 
Figure~\ref{Fig3}(a) shows the experimentally determined $|\chi^{(3)}|$ as a function of temperature for different driving frequencies of $f=0.4$, 0.5, and 0.7~THz.
While at a fixed temperature the absolute value of $|\chi^{(3)}|$ increases at lower frequencies, for all three frequencies $|\chi^{(3)}|$ monotonically decreases at elevated temperature and vanishes above 80~K. 
By normalizing $|\chi^{(3)}|$ with respect to the maximum value of each frequency [see Fig.~\ref{Fig3}(b)], the three curves overlap well with each other, exhibiting a frequency-independent temperature dependence of nonlinear response. 

This master curve reveals a characteristic temperature scale, below which a well-defined electronlike quasiparticle band is formed right below the Fermi energy, as directly observed by angle-resolved photoemission spectroscopy \cite{Liu2018}.
In comparison with density functional theory calculations, this quasiparticle band is strongly renormalized with an effective mass of 13.5$m_e$, indicating a strong correlation effect.
For the spatially more extended $4d$ orbitals in CaRuO$_3$, the strong correlations result from the Hund's coupling \cite{Dang2015,Dang2015a},
which has been found also to account for a similar temperature dependent evolution of heavy quasiparticles in a related compound Sr$_2$RuO$_4$ \cite{Mravlje11,Shen07}.
The characteristic temperature is much higher than the scale, i.e. $T<1.5$~K, where a Fermi-liquid type $T^2$ dependence of electrical resistivity appears \cite{Schneider2014}, but reflects a coherence-incoherence crossover of the heavy quasiparticles related to the strong correlations \cite{Mravlje11}.
At higher temperatures a good metallic response in optical conductivity is absent as manifested by the disappearance of Drude-like behavior at low frequencies \cite{Geiger2016}. 

It has been established that THz high-harmonic radiation can be generated from nonlinear kinetics of relativistic quasiparticles with linear dispersion relation (see e.g. \cite{Kovalev2020,Dantas21,Germanskiy22}). According to the band-structure calculations \cite{Dang2015a,Schneider2014}, some of the bands close to the Fermi surface in CaRuO$_3$ may likely follow a linear dispersion relation, which potentially also leads to THz third-harmonic generation. However, these bands have not been experimentally confirmed so far. Moreover, the harmonic generation associated with a linear dispersion should be well visible already at room temperature.
Therefore, our observed enhancement of the third-harmonic generation at low temperatures is unlikely governed by these bands. In contrast, we ascribe the observed nonlinear response to THz field driven kinetics of the experimentally observed heavy quasiparticles in CaRuO$_3$ with an energy-dependent scattering rate.

We describe the field driven transport by using the time-dependent Boltzmann equation
\begin{align}
\label{boltzmann}
\left(\frac{\partial}{\partial t}+\frac{1}{\tau}\right)f(t,\textbf{p})-e\textbf{E}(t)\cdot \nabla_{\textbf{p}} f(t,\textbf{p}) = \frac{f_0(\textbf{p})}{\tau},
\end{align}
where $f(t,\textbf{p})$ denotes the time-dependent distribution function at time $t$ and momentum $\textbf{p}$ for a band that is characterized by a dispersion relation $\epsilon(\textbf{p})$.
$\textbf{E}(t)$ is the electric field of the THz drive and $f_0(\textbf{p})$ is the Fermi-Dirac distribution at the equilibrium state. 
For a parabolic band with an energy-independent relaxation time in a slowly varying electric field, Eq.~(\ref{boltzmann}) reduces to the Drude model and the corresponding transport behavior follows the Ohm law, i.e. $\textbf{j}=\sigma \textbf{E}$.
In the relevant frequency range of our experiment, previous studies in the linear response regime have revealed clear deviation of the transport behavior in CaRuO$_3$ from the description by the Drude model \cite{Lee2002,Kamal2006,Schneider2010,Geiger2016,Wang2021}, and also predicted energy-dependent relaxation time due to electron-electron correlations \cite{Dang2015,Dang2015a}.
To parametrize the experimentally observed energy-dependent relaxation time in 
Ref.~\cite{Schneider2014}, we assume a linear dependence in our simulation, i.e. $\tau(\textbf{p})=\tau_0\left(1+\alpha\textbf{p}^2\right)$
with $\tau_0$ being the energy-independent scattering time and $\alpha < 0$ describing the increased scattering rate at higher energy.

We solve Eq.~(\ref{boltzmann}) numerically for a heavy electron parabolic band with the experimental parameters \cite{Schneider2014,Liu2018}, i.e. an enhanced quasiparticle mass of $13.5 m_e$, a Fermi energy of $E_F= 7$~meV, $T=300$~K, $\tau_0=1$~ps and a THz peak electric field of $E_{\text{max}}=50$~kV/cm.
The experimental values of $\alpha$ at various temperatures are given in Fig.~\ref{Fig3}(c) \cite{Schneider2014}. 
In the simulation a multicycle THz pulse with a Gaussian envelop is adopted, whose central frequency and bandwidth are set in accord with the experimental values.
The THz field driven current density is evaluated by
$\mathbf{j}(t) = -e\int \frac{d\textbf{p}^3}{(2\pi \hbar)^3}f(t,\textbf{p})\nabla_{\textbf{p}}\epsilon(\textbf{p})$, thereby we obtain the time-dependent emitted THz electric field as the time derivative of the current density.

The emitted THz signal contains components of the fundamental frequency and the third harmonic radiation.
Through Fourier transformation we derive the third-order nonlinear susceptibility as $|\chi^{(3)}| \propto \frac{E_{3f}}{E_f^3}$ with $E_{3f}$ and $E_f$ being integrated amplitudes in the frequency domain around $3f$ and $f$, respectively, which as a function of temperature is presented in Fig.~\ref{Fig3}(b).
One can see that corresponding to a negligible energy dependence of the relaxation time (i.e. $\alpha \approx 0$) at high temperatures [see Fig.~\ref{Fig3}(c)], the third-order nonlinear susceptibility is essentially zero. 
This means that a heavier effective mass alone (corresponding to a weaker curvature of the parabolic dispersion) cannot account for the experimentally observed THG. 
With decreasing temperature, the absolute value of $\alpha$ increases [Fig.~\ref{Fig3}(e)], which leads to an enhancement of the THG [Fig.~\ref{Fig3}(c)]. Therefore, a more evident dependence of the scattering rate on energy is responsible for the enhanced THz nonlinearity.    

To further illustrate this effect, in Fig.~\ref{Fig3}(d) we compare the amplitude of emitted THz field for zero and a negative value of $\alpha$ at 5~K.
While for $\alpha=0$ only radiation of the fundamental frequency is emitted, for $\alpha < 0$ higher-order harmonic generation becomes very evident.
Microscopically, the field-driven nonlinear current density is closely related to the deviation of the charge-carrier distribution from the Fermi-Dirac function.
The snapshot of the distribution corresponding to the peak THz field (i.e. $E=E_\text{max}$) is presented in Fig.~\ref{Fig3}(e) together with the zero-field distribution at 5~K for comparison.
Under the drive of the THz field, the distribution is clearly stretched, which cannot be described by a Fermi-Dirac function, featuring the important contribution of the nonthermal states to the nonlinear response.

Even for $\alpha=0$ the strong THz electric field can drive the system far from thermal equilibrium, therefore it is instructive to quantize the corresponding nonthermal effects due to the energy-dependent scattering.
As displayed in Fig.~\ref{Fig3}(f), we evaluate the changes of distribution function at the peak field $E_\text{max}$ for $\alpha$'s at different experimental temperatures with respect to that of $\alpha=0$.
The changes are more significant at lower temperatures corresponding to larger absolute values of $\alpha$ [see Fig.~\ref{Fig3}(c)].
As reflected by the positive and negative changes around the Fermi level, the charge carrier distribution is more strongly stretched away from a Fermi-Dirac function with increasing $\alpha$, which lead to the observed enhancement of THz harmonic generation. 

To conclude, by driving the strongly correlated metal CaRuO$_3$ with intense terahertz field, we observed third-harmonic radiation below 80~K, in agreement with the temperature dependence of a heavy quasiparticle band emerging close to the Fermi surface. 
The field-driven kinetics of the heavy quasiparticles is simulated by adopting a Boltzmann transport equation with an energy-dependent scattering rate, which reflects the previously observed non-Drude behavior. 
The observed third-harmonic generation is not necessarily a unique feature for CaRuO$_3$, but should represent in general a peculiar nonlinear characteristic for nonequilibrium states in strongly correlated metals close to a quantum critical point \cite{Dalidovich04,Green05,Oka05,Mitra06,Hogen2008,
Shao22,Murakami22,Prochaska20,Yang2023}.
We anticipate that our work will motivate further experimental and theoretical studies to investigate the universal characteristic nonlinear responses in quantum critical metals.

\begin{acknowledgments}
We thank R. Dantas, Yang Liu, J. Mravlje, and M. Scheffler for very helpful discussions and P. Pilch for assistance in figure preparation.
The work in Cologne was partially supported by the DFG  via  Project No. 277146847 — Collaborative Research Center 1238: Control and Dynamics of Quantum Materials (Subproject No. B05).
The work in Augsburg was supported by the DFG through
Project No. 492547816 - TRR 360:  Constrained Quantum Matter.
Se. E. acknowledges current support by the Japan Society for the Promotion of Science (JSPS) through KAKENHI Grant No. 23H05431 and by the Marubun foundation through their Exchange Grant.
Z.W. acknowledges support by the European Research Council (ERC) under the Horizon 2020 research and innovation programme, grant agreement No. 950560 (DynaQuanta).
\end{acknowledgments}

\bibliographystyle{apsrev4-2}
\bibliography{CaRuO3_bib}

\end{document}